# Systemic Methodological Dysfunction in Statistical Research for Clinical Decisions

Charles F. Manski

Department of Economics and Institute for Policy Research
Northwestern University, Evanston, IL 60208 USA
cfmanski@northwestern.edu

July 2026

Abstract

I critique a set of entrenched methodological conventions that collectively create systemic dysfunction in statistical research for clinical decisions. These include: (1) the prevalent use of hypothesis tests to compare treatments, (2) remoteness from patient care of the methods used to evaluate the accuracy of predictions of patient outcomes, (3) poor practice of meta-analysis to combine findings across studies, and (4) widespread research with incredible certitude. It appears that the dysfunction is held in place by three factors: (i) rudimentary instruction in statistical methodology received by medical students and residents, (ii) reliance of clinical researchers on consulting biostatisticians, wo act as statistical gatekeepers in evaluation of grant proposals and paper submissions, and (iii) institutional practices of research funding agencies, medical journals, and governmental bodies that regulate medical treatment. I conjecture that systemic changes are necessary to break the existing impasse, moving statistical research to a better equilibrium.

.

Acknowledgements: I am grateful to Jeff Dominitz, John Mullahy, and Filip Obradovic for helpful comments.

## 1. Introduction

I write with heavy concern about the practice of statistical research for clinical decisions. My perspective has formed gradually. I am an econometrician with no formal training in medicine. As such, I should not and do not give advice regarding patient care. Yet I feel qualified to contribute usefully to the methodology of evidence-based medicine. This matter lies within the expertise of econometricians, statisticians, and decision analysts.

Clinical research seeks to predict patient outcomes conditional on their observed covariates and potential treatments. The aim is to inform clinical decisions. Development of methodology for prediction to inform decision making has long been a core objective of econometrics. I have discussed aspects of the history in Manski (2007, 2021a).

Around 2000, I began to use patient care to illustrate broad methodological issues in prediction and decision making. Moving beyond illustration, I have since performed new methodological research addressing substantive medical subjects. As my experience has grown, my perspective on statistical research for clinical decisions has become increasingly negative. When I was an outsider entering the field, I optimistically supposed that evidence-based medicine must be healthy overall. I perceived that the field could benefit from incorporation of some concepts developed in econometrics and statistical decision theory. I anticipated that clinical researchers would welcome these concepts and would use them once they become aware of them.

Over time, I have reluctantly concluded that my optimism was naïve. I now see evidence-based medicine as a deeply ill enterprise. The central problem is adherence to a host of ill-motivated, entrenched methodological conventions. I initially thought that these conventions were isolated problems that could be addressed one at a time. I have concluded that they collectively manifest a systemic dysfunction in clinical research.

As I see it, the dysfunction is held in place jointly by three factors. One is the rudimentary, often rote, instruction in statistical methodology received by medical students and residents. I appreciate that the intense four-year experience of medical school lacks time to train students to be competent econometricians or statisticians. However, the limited time now devoted to study of methodology should not indoctrinate students to apply misguided statistical practices in cookbook fashion. I see considerable scope for deeper instruction post receipt of the MD degree, when fledging physicians undertake residencies and research fellowships.[1]

The second factor, which stems partly from the instructional deficit, is the reliance of clinical researchers on the guidance of consulting biostatisticians. These individuals, who commonly hold faculty positions in medical schools, act as statistical gatekeepers in evaluation of grant proposals and paper submissions. They approve methodological practices that adhere to convention. They reject others for the sin of violating the conventions.

The third factor is the institutional buy-in of research funding agencies, medical journals, and governmental bodies that regulate medical treatment. These institutions make adherence to convention almost a prerequisite for award of grants, acceptance of paper submissions, and approval of treatment innovations. Clinical researchers cannot have successful careers if they are unable to obtain research funding or publish their work in high-impact journals. Developers of new drugs and medical devices cannot improve patient care unless their innovations are approved by the US FDA or similar agencies in other countries.

This paper fleshes out why I perceive systemic methodological dysfunction in the statistical aspects of clinical research. My focus throughout is the use of research to inform clinical decisions, not research performed to advance medical science per se. Hence, I find it useful to begin in Section 2 with specification of a realistic yet simple class of decision problems that often arise in patient care. In these decisions, a clinician must choose between two treatments for a possible disease. The expected effect of each treatment on patient welfare depends on the patient's personalized probability of illness; that is, conditional on observed patient covariates. I use this class of decisions to frame discussion in the remainder of the paper.

[1] It has become common for medical students and physician researchers, aiming to broaden their backgrounds beyond clinical medicine, to obtain degrees in Master of Public Health programs. Unfortunately, these programs typically offer only routine instruction in statistical conventions.

Sections 3 through 6 describe and critique a set of entrenched methodological practices that severely diminish the usefulness of clinical research to patient care. Section 3 addresses the prevalent use of hypothesis tests and confidence interval coverage to compare treatments. Section 4 considers approaches that clinical researchers use to evaluate the accuracy of illness predictions. Section 5 critiques the conventional practice of meta-analysis to combine findings across research studies. Section 6 cautions against multiple conventional practices of research with incredible certitude. The concluding Section 7 conjectures systemic changes that could break the existing impasse, moving statistical practice to a better equilibrium.

*Background*

Given that this paper provides a personal perspective, I think it appropriate before proceeding to summarize the background underlying my severe critique. My research on methodology for clinical research has been reported in articles in journals in econometrics, statistics, and epidemiology that publish new research methodology communicated in a rigorous mathematical manner. My work has not been published in medical journals, which lack these qualities. Indeed, medical journals manifest pronounced algebra aversion. Recognizing that clinical researchers and clinicians do not ordinarily read methodological journals, I have sought to communicate verbally to these communities, particularly though my expository book *Patient Care under Uncertainty* (Manski, 2019a).

A persistent and continuing subject of my work has been the design and analysis of randomized clinical trials to improve how they inform clinical decisions (Manski, 2004, 2019b, 2024, 2025; Manski and Tetenov, 2016, 2019, 2021). I have sharply criticized the longstanding hegemony of hypothesis testing as a framework to guide choice of sample size and interpretation of trial data. I have recommended its replacement by the statistical decision theory pioneered by Wald (1939, 1950). I have specifically recommended use of maximum regret to measure the informativeness of trials.

Another persistent and continuing subject of study has been prediction of patient outcomes when identification problems impede credible inference and decision making. Statistical theory characterizes the inferences that can be drawn by observing a finite data sample drawn from a population. Identification analysis studies inferential difficulties that persist even when sample size grows without bound. A population feature (aka parameter) of interest is point-identified if empirical knowledge of aspects of the population and maintained assumptions reveal it precisely. The feature is partially identified if the available information reveals that it lies in an informative set larger than a single point. From the 1990s onward, partial identification analysis has become widespread in econometrics and has more recently spread to epidemiology. Yet it remains essentially non-existent in clinical research reported in medical journals. Mullahy *et al.* (2021) sought to bring the subject to the attention of clinical researchers.

Clinical settings with credible partial identification, several of which will be described later in this paper, arise using either trial or observational data. Horowitz and Manski (2000) and Scharfstein *et al.* (2004) studied partial identification of illness probabilities when data on patient outcomes and/or covariates may be missing nonrandomly. Manski (2013, 2021b) and Cassidy and Manski (2019) addressed the performance and interpretation of diagnostic tests. Manski, Tambur, and Gmeiner (2019) investigated prediction of kidney transplant outcomes with partial knowledge of HLA mismatch. Manski (2020a) critiqued the conventional performance of meta-analysis to aggregate prediction findings across disparate studies. Manski (2018, 2023) and Li, Litvin, and Manski (2023) examined the problem of learning about personalized treatment response when medical journals report only treatment response for large subgroups. Manski (2025) studied dosage choice for cancer drugs when clinical trials provide only limited evidence on dose response.

I have complemented these specific contributions by writing on some broad methodological themes in clinical research and practice. Manski (2019a) devoted considerable attention to the role that clinical practice guidelines play in clinical decisions and to the FDA drug approval process. Manski (2022) and Manski, Mullahy, and Venkataramani (2023) addressed the controversial issue of using race as a patient covariate when predicting health outcomes.

Beyond performing methodological research, I have learned much about norms in the practice of clinical research through professional interactions with clinical researchers. Most intensive has been my participation in 2020-2026 as Multiple Principal Investigator in a collaborative project (US NIH Grant R01DK131164) studying the trajectory and cost of treatment of liver cirrhosis (e.g., Obradovic *et al.*, 2025). I have also benefited from instructive continuing interactions with oncologists who study dose response to cancer drugs (e.g., Earl *et al.*, 2026).

## 2. Binary Treatment Choice to Maximize Patient Expected Welfare

A stream of research on clinical decisions has addressed choice between two treatments from the perspective of maximization of patient expected welfare. Uncertainty arises because a clinician must choose a treatment with incomplete knowledge of treatment response. In a canonical version of the decision problem, the clinician does not know whether a patient is ill but does know the probability of illness conditional on observed patient covariates. Early formalizations of this problem of covariate-personalized treatment include Pauker and Kassirer (1975, 1980) and Phelps and Mushlin (1988).

Among decision problems of this type, I have found it particularly interesting to study choice between surveillance and aggressive treatment of patients at risk of potential disease (e.g., Manski, 2018, 2019a). For example, surveillance of women at risk of breast cancer may mean undergoing periodic mammograms and clinical exams, while aggressive treatment may mean preventive drug treatment or even mastectomy. Other familiar examples are choice between surveillance and drug treatment for patients at risk of heart disease or diabetes. Yet others are choice between surveillance and adjuvant drug treatment of patients who have been treated for localized cancer and are at risk of metastasis. A semantically distinct but logically equivalent decision is choice between diagnosis of patients as healthy or ill. With diagnosis, the concern is not to judge whether a patient will develop a disease in the future but whether the patient is currently ill and requires treatment. These decisions are common and familiar to clinicians and patients alike.

## 2.1. The Decision Problem with Known Illness Probabilities and Patient Preferences

To understand the decision problem fully, it is necessary to express it mathematically. I do so here. I repeatedly use this formalization later in the paper.

Let treatment $t = A$ denote surveillance and $t = B$ denote aggressive treatment. Let $y(A)$ and $y(B)$ be potential binary illness outcomes, with $y = 1$ denoting that the patient will develop the disease of concern and $y = 0$ otherwise. Let $x$ be patient covariates observed by a clinician. Let $P_x(t) \equiv P[y(t) = 1|x]$ be the probability that a patient with covariates $x$ will develop the disease if the patient receives treatment $t$.

Patient welfare with each care option depends on whether a patient will or will not develop the disease. Let $u_x(y, t)$ denote the welfare of treatment $t$ to a patient with covariates $x$ should disease outcome $y$ occur. Consider a clinician who chooses a care option without knowing the disease outcome but with knowledge of illness probabilities.

The expected welfare of each treatment $t$ is

$$(1) \qquad P_x(t)\cdot u_x(1, t) + [1 - P_x(t)]\cdot u_x(0, t).$$

Maximization of expected welfare yields the optimal treatment rule

(2a) Choose A if

$$P_x(A)\cdot u_x(1, A) + [1 - P_x(A)]\cdot u_x(0, A) \geq P_x(B)\cdot u_x(1, B) + [1 - P_x(B)]\cdot u_x(0, B),$$

(2b) Choose B if

$$P_x(B)\cdot u_x(1, B) + [1 - P_x(B)]\cdot u_x(0, B) \geq P_x(A)\cdot u_x(1, A) + [1 - P_x(A)]\cdot u_x(0, A).$$

A clinician who knows patient welfare and illness probabilities chooses the optimal treatment, yielding expected welfare $\max\{P_x(A)\cdot u_x(1, A) + [1 - P_x(A)]\cdot u_x(0, A), P_x(B)\cdot u_x(1, B) + [1 - P_x(B)]\cdot u_x(0, B)\}$.

Decision rule (2) is simple, but it is instructive to call attention to further simplifications that occur when treatment operates on disease in different polar ways. In some clinical settings, aggressive treatment always prevents disease development whereas surveillance does not; thus, $P_x(B) = 0$ and $P_x(A) > 0$. In other settings, treatment does not affect disease development but may affect the severity of illness when it occurs; thus, $P_x(B) = P_x(A)$, but $u_x(1, A)$ may differ from $u_x(1, B)$. In settings of both types, calculation of threshold probabilities of disease suffices to determine the optimal treatment, as shown below.

Suppose that $P_x(B) = 0$. Then (2) reduces to

(3a) Choose A if $P_x(A)\cdot u_x(1, A) + [1 - P_x(A)]\cdot u_x(0, A) \geq u_x(0, B)$,

(3b) Choose B if $u_x(0, B) \geq P_x(A)\cdot u_x(1, A) + [1 - P_x(A)]\cdot u_x(0, A)$.

Hence, the optimal treatment depends on the magnitude of $P_x(A)$ relative to the threshold value that equalizes the expected welfare of the two treatments:

$$(4) \qquad P^*_x(A) \equiv \frac{u_x(0, A) - u_x(0, B)}{u_x(0, A) - u_x(1, A)}.$$

It is generally reasonable to expect that surveillance yields higher expected welfare when a patient will remain healthy rather than develop the disease; that is, $u_x(0, A) > u_x(1, A)$. Then surveillance is optimal when $P_x(A) \leq P^*_x(A)$ and aggressive treatment is optimal when $P_x(A) \geq P^*_x(A)$.

Now suppose that $P_x(A) = P_x(B) \equiv P_x$. Then (2) reduces to

(5a) Choose A if $P_x\cdot u_x(1, A) + (1 - P_x)\cdot u_x(0, A) \geq P_x\cdot u_x(1, B) + (1 - P_x)\cdot u_x(0, B)$,

(5b) Choose B if $P_x\cdot u_x(1, B) + (1 - P_x)\cdot u_x(0, B) \geq P_x\cdot u_x(1, A) + (1 - P_x)\cdot u_w(0, A)$.

Now the optimal treatment depends on the magnitude of $P_x$ relative to the threshold value that equalizes the expected welfare of the two treatments:

$$(6) \qquad P^*_x = \frac{u_x(0, A) - u_x(0, B)}{[u_x(0, A) - u_x(0, B)] + [u_x(1, B) - u_x(1, A)]}.$$

It often is reasonable to suppose that surveillance yields higher expected welfare when a patient will not develop the disease and that aggressive treatment yields higher welfare when a patient will develop the disease; that is, $u_x(0, A) > u_x(0, B)$, and $u_x(1, B) > u_x(1, A)$. Then $0 < P^*_x < 1$. Treatment A is optimal if $P_x \leq P^*_x$ and B is optimal if $P_x \geq P^*_x$.

### 2.2. Objectives of Clinical Research

To inform the above decision problem, the objective of clinical research should be to learn the illness probabilities and patient welfare that determine expected welfares. These two objectives have historically been addressed separately. Clinical researchers and biostatisticians perform the former research. Health economists perform the latter.

My concern in this paper is with statistical research aiming to learn illness probabilities. Economic research aiming to learn patient welfare has also been afflicted by over-adherence to certain conventions, but these issues are distinct and will not be addressed here; see the discussion in Manski and Mullahy (2026). In what follows, I assume that patient welfare is known and I focus on the problem of learning illness probabilities.

Observe that optimal binary treatment choice does not require precise knowledge of illness probabilities. It only requires determination of the correct inequality in equation (2). The criterion for optimality is particularly simple in choice between surveillance and aggressive treatment when either (i) aggressive treatment always prevents disease development or (ii) treatment does not affect disease development but affects the severity of illness when it occurs. In these settings, all that is necessary to optimize treatment is to determine whether $P^*_x(A)$ or $P^*_x$ is above or below the threshold in equation (4) or (6) respectively.

A clinician who chooses treatment B when A yields larger expected welfare incurs a Type A error. The patient suffers a loss in expected welfare, commonly termed regret, that equals the difference between the expected welfare of A and B. Analogously, a clinician who chooses A when B is superior incurs a Type B error and the patient likewise suffers regret.

## 3. Using Hypothesis Tests and Confidence Interval Coverage to Compare Treatments

### 3.1. Hypothesis Tests

Medical research has long used classical hypothesis testing to compare the mean patient outcomes of two treatments. The standard practice has been to view treatment A as the status quo (aka *standard care*) and B as an innovation. The usual null hypothesis is that the mean outcome with B is no better than that with A. The alternative is that B is superior to A. If the null is not rejected, it is recommended that A continue to be the standard treatment. If the null is rejected, with B appearing better than A, it is recommended that B become the standard treatment. This type of test is performed routinely in academic research. It is also entrenched in FDA drug approval, with the distinction that the recommendation is whether to approve the drug for marketing.[2]

If outcomes are defined as patient expected welfare as in Section 2, the null hypothesis is that weak inequality (2a) holds and the alternative is that (2b) holds as a strict inequality. However, medical applications of hypothesis testing rarely make explicit reference to patient welfare. Instead, clinical researchers designing trials pre-declare that the trial will evaluate some primary outcome (aka primary endpoint) and various secondary outcomes. The convention has been to test a null hypothesis concerning the primary outcome. Secondary outcomes are often evaluated less formally. A commonly designated primary outcome is a probability of illness, and a common null hypothesis is that the probability of illness with treatment A is less than or equal to that with B; that is, $P_x(A) \leq P_x(B)$. The alternative is $P_x(A) > P_x(B)$. In terms of the model of Section 2, this is the special case where $u_x(1, A) = u_x(1, B) = 0$ and $u_x(0, A) = u_x(0, B) = 1$.

---

[2] Indeed, hypothesis testing is so deeply embedded in clinical research that journal articles regularly report test findings, p-values, and confidence intervals even using administrative or other observational data not generated by a random sampling process as assumed in the theory of testing.

The convention in testing has been to fix the probability of rejecting the null hypothesis when it is correct, a Type I (or Type A) error, at 0.05. Then sample size determines the probability of rejecting the alternative hypothesis when it is correct, a Type II (or Type B) error. Statistical power is one minus the probability of a Type II error. The convention in designing randomized trials has been to choose a sample size that yields specified power, typically 0.08 to 0.9, at a value of the effect size that is deemed clinically important. There is no consensus on the appropriate value of a clinically important effect size.

Manski and Tetenov (2016) observed that there are several reasons why hypothesis testing does not provide an appropriate criterion for clinical decisions. These include

*1. Use of Asymmetric Error Probabilities*: The convention has been to set the probability of Type I error at 5% and that of Type II error at 10-20%. The theory of hypothesis testing gives no rationale to select these error probabilities. It gives no reason why a clinician should tolerate a substantially greater probability of Type II than Type I error.

*2. Inattention to Magnitudes of Patient Welfare Losses When Errors Occur*: A clinician should care about the magnitudes of the losses to patient welfare associated with errors. A given error probability should be less acceptable when the welfare difference between treatments is larger. Hypothesis testing does not take this into account.

3. *Limitation to Settings with Two Treatments*: A clinician often chooses among several treatments and many clinical trials compare more than two treatments. Yet the standard theory of hypothesis testing only contemplates choice between two treatments. Statisticians have struggled to extend it to deal sensibly with comparisons of multiple treatments.

Further critique has been presented in the *American Statistical Association Statement on Statistical Significance and P*-Values (Wasserstein and Lazar, 2016). Two principles of the Statement are:

"Scientific conclusions and business or policy decisions should not be based only on whether a *p*-value passes a specific threshold."

"A *p*-value, or statistical significance, does not measure the size of an effect or the importance of a result."

Aiming to avoid publication of statistically insignificant results ex ante, researchers often perform studies and report findings only for groups whose sample sizes are large enough to perform tests with conventional Type I and II error probabilities. Moreover, researchers sometimes selectively report findings that are statistically significant ex post by standard criteria. This reporting practice has been recognized to generate publication bias (Ioannidis, 2005; Wasserstein and Lazar, 2016).

Bayesian researchers have long criticized the use of hypothesis testing to design trials and make treatment decisions. The literature on Bayesian statistical inference rejects the frequentist foundations of hypothesis testing, arguing for superiority of the Bayesian practice of using sample data to transform a subjective prior distribution on treatment response into a subjective posterior distribution (e.g., Speigelhalter *et al.,* 1994). As I see it, Bayesian inference is a logically coherent approach to learning that is well-motivated when a prior distribution has a credible foundation. Berger (1985) phrased this basic issue well when he wrote (p. 121): "A Bayesian analysis may be 'rational' in the weak axiomatic sense, yet be terrible in a practical sense if an inappropriate prior distribution is used."

3.2. Confidence Interval Coverage

Closely related to the use of hypothesis tests to compare mean treatment outcomes is the common research practice of examining the coverage of a confidence interval for the average treatment effect (ATE); that is, the difference between the mean outcomes for treatments B and A.[3] When a computed confidence interval for the ATE covers the value zero, clinical researchers often conclude that treatment B is not superior to treatment A. The rationale appears to be that inclusion of the value zero in a confidence interval is equivalent to the statement that some test of the null hypothesis [ATE = 0] does not reject the null.[4]

Statisticians and econometricians grounded in the theory of hypothesis testing and confidence intervals understand that failure to reject a null hypothesis does not imply acceptance of the hypothesis. They know that, for every value c contained in a confidence interval, some test of the null hypothesis [ATE = c] does not reject this null. The value c = 0 has no special status in the theory. Statisticians and econometricians emphasize these basic points when teaching statistical theory.

Nevertheless, clinical research often makes the elementary mistake of interpreting the event that 0 lies in a confidence interval for the ATE to be evidence that ATE = 0. This interpretation has no foundation. Moreover, it can yield troubling consequences for clinical decisions. I illustrate below, drawing on Manski and Tetenov (2021).

*Interpreting the Findings of an Early Trial of Treatments for COVID-19*

At the beginning of the COVID-19 pandemic, Cao *et al.* (2020) reported on a randomized trial in China comparing standard-care treatment of severe cases of COVID-19 with standard-care combined with the drug pair lopinavir–ritonavir. The trial assigned 99 hospitalized adult patients to the lopinavir–ritonavir group and 100 to the standard-care only group. The pre-declared primary outcome measured time to clinical improvement. A secondary outcome was mortality within 28 days.

---

[3] Focus on the ATE rather than another parameter to measure distributional differences in outcomes across treatment is itself an entrenched convention in clinical research. Mullahy (2018) discusses alternative measures.

[4] A further convention is computation of a confidence interval with 95% coverage probability. This convention has no more foundation than the prevailing view in hypothesis testing that a Type I error should occur with probability 0.05.

The authors summarized the primary finding as follows (p. 1): "In a modified intention-to-treat analysis, lopinavir–ritonavir led to a median time to clinical improvement that was shorter by 1 day than that observed with standard care (hazard ratio, 1.39; 95% CI, 1.00 to 1.91)." Regarding mortality, 19 of the 99 patients assigned to lopinavir-ritonavir died within 28 days and 25 of the 100 receiving only standard care died. The authors characterized this finding as follows (p. 1): "Mortality at 28 days was similar in the lopinavir–ritonavir group and the standard-care group (19.2% vs. 25.0%; difference, −5.8 percentage points; 95% CI, −17.3 to 5.7)." They concluded (p. 1): "In hospitalized adult patients with severe COVID-19, no benefit was observed with lopinavir-ritonavir treatment beyond standard care."

A clinician might reasonably view the estimated reductions in median time to clinical improvement and in mortality to be suggestive, albeit not definitive, evidence that treatment with lopinavir–ritonavir was beneficial relative to standard care alone. Yet the study authors concluded (p. 1): "no benefit was observed with lopinavir–ritonavir treatment beyond standard care." This conclusion was reached because estimated treatment effects were not statistically significant, having 95% confidence intervals that cover zero. Subsequently, COVID-19 treatment guidelines issued by National Institute of Health (2020) cited the absence of statistical significance when it characterized the study as having negative findings.

Requiring statistical significance to recommend a treatment innovation shows deference to standard care, placing the burden of proof on the innovation. A clinician might perhaps argue that it is reasonable to place the burden on an innovation when standard care is known to yield good patient outcomes. However, this argument lacked credibility in the COVID-19 setting. In early 2020, standard care for this new disease evolved rapidly in the face of great uncertainty. At the time, there did not yet exist a standard treatment yielding notably good patient outcomes.

## 3.3. Primary Outcomes, Secondary Outcomes, and Expected Patient Welfare

I mentioned in Section 3.1 the conventional practice in design and analysis of trials to separately evaluate a designated primary outcome and various secondary outcomes, with focus on the primary outcome. I noted that clinical research rarely tests hypotheses regarding expected patient welfare, which should be the estimand of interest for clinical decisions.

Welfare assessment combines the multiple outcomes of a treatment to determine its net potential impact on a patient. It recognizes that treatment choice may require making tradeoffs: One treatment may be better for a patient in some respects but worse in others. Clinicians are aware of this qualitatively when they discuss the efficacy and adverse effects (aka toxicity or side effects) of drug and surgical treatments. Yet conventional statistical analysis does not seek to evaluate the combined impacts of efficacy and adverse effects on patient welfare. Instead, the prevalent practice views efficacy as the primary outcome and adverse effects as secondary outcomes. The FDA provided its perspective on primary and secondary outcomes in U.S. Food and Drug Administration (2022).

Making a sharp distinction made between primary and secondary outcomes is reasonable from a welfare perspective when the primary endpoint is the dominant determinant of patient welfare or, put another way, when there is little variation in secondary outcomes across treatments. It is not reasonable otherwise. I provided an illustration in Manski (2019a), summarized below.

*Illustration*: Consider trials assessing anti-coagulation drugs such as warfarin as treatments for atrial fibrillation. It has been standard to designate the occurrence of stroke as the primary endpoint, with major bleeding viewed as a secondary outcome; see, for example, Ezekowitz *et al.* (1992). However, the risk of serious bleeding with warfarin treatment has been thought sufficiently serious as to prompt strong warnings to patients. From the perspective of patient welfare, it may be more appropriate to view occurrence of strokes and major bleeding as jointly relevant outcomes rather than to consider the former to be the primary outcome and the latter to be secondary. Thus, we may view a patient receiving warfarin as experiencing one of four possible outcomes: (1) neither stroke nor major bleeding, (2) stroke and no major bleeding, (3) major bleeding and no stroke, (4) both stroke and major bleeding. Which outcome will occur is not known at the time of treatment. The standard recommendation of medical economics would be to determine the probability that each outcome will occur, multiply this probability by patient utility in the event of this outcome, and use expected utility to measure patient welfare.

### 3.4. Treatment Choice using Statistical Decision Theory

In Manski (2004, 2019a, 2021, 2024) and Manski and Tetenov (2016, 2019, 2021), I have argued that the seminal Wald (1939, 1950) development of statistical decision theory provides a coherent framework for use of sample data on treatment response to choose treatments. I explain here.

A standard formalization of decision making under uncertainty considers a decision maker who chooses among a set of feasible actions. The outcome achieved by an action depends on an unknown feature of the environment, often called the *state of nature*. The decision maker wants to maximize some outcome, but this is unachievable with certainty because the true state of nature is unknown. The decision maker lists all states that he believes could occur. This list, called the *state space*, expresses partial knowledge.

In a medical setting, the decision maker is a clinician. The actions are a set of feasible treatments. In the setting of Section 2, the clinician wants to maximize patient expected welfare. A state of nature is a pair of feasible values for illness probabilities $P_x(A)$ and $P_x(B)$.

Wald considered the use of sample data to make decisions. He posed the task as choice of a *statistical decision function*, which maps potentially available data into a choice among the feasible actions. When considering treatment choice, a statistical decision function is sometimes called a *statistical treatment rule*. I use this term henceforth.

Wald proposed that a decision maker evaluate a statistical treatment rule by the mean outcome it yields across realizations of the sampling process; thus, the theory is frequentist. He studied a three-step process. One first specifies the feasible actions, the outcome function, and the state space. One next eliminates inadmissible decision functions---a decision function is inadmissible if there is another that yields at least as good mean sampling performance in every state and strictly better mean performance in some state. Finally, one uses some reasonable criterion to choose an admissible decision function.

3.4.1. Decision Criteria

What are reasonable ways to choose an admissible decision function? Wald focused on the *maximin* criterion and on maximization of subjective mean welfare, which is often called Bayesian decision making.[5] Savage (1951) proposed *minimax regret (MMR)*.

Bayesian decision making may be compelling given a credible prior distribution on the state space. However, Bayesians have long struggled to provide guidance on credible priors. The matter continues to be controversial. See the disparate views regarding Bayesian analysis of randomized trials conveyed by the authors and discussants of Spiegelhalter *et al.* (1994).

Wald argued that a reasonable way to act without specifying a prior distribution is to use a decision criterion that achieves uniformly satisfactory results, whatever the true state of nature may be. There are various ways to formalize this idea. The two most prominent are the maximin and minimax-regret (MMR) criteria.

The maximin criterion is intentionally ultra-conservative. It chooses a treatment that maximizes the minimum outcome that might possibly occur across all feasible states of nature. The MMR criterion considers each state of nature and computes the incremental loss that would occur if one were to choose a specified treatment, rather than one that maximizes the outcome of interest in this state. This incremental loss, called *regret*, measures the nearness to optimality of the specified action in the specified state. The decision maker must choose without knowing the true state. To achieve a uniformly satisfactory result, he computes the maximum regret of each action---the maximum distance from optimality that the action would yield across all possible states. The MMR criterion makes a choice that minimizes maximum regret.

[5] It has been common to think of Bayesian transformation of a prior into a posterior distribution as antithetical to frequentist statistics. However, Wald provided a frequentist perspective on Bayes decisions. It can be shown that maximization of a subjective mean outcome, which is a frequentist decision criterion, ordinarily yields the same decisions as would occur if one were to perform Bayesian inference, wherein one uses a prior distribution and data to form a posterior distribution and then maximizes the posterior mean outcome.

Analysis has shown that the MMR criterion behaves well in the context of treatment choice with data from a randomized trial. In common settings with trial data on outcomes that take a bounded range of values, it has been found that the MMR rule is well approximated by the *empirical success* (ES) rule, which chooses the treatment with the highest average outcome in the trial. The ES rule provides a simple and effective way to use trial findings.

Maximum regret quantifies how incomplete knowledge of the true state of nature diminishes the quality of decisions. A treatment rule with small maximum regret is uniformly near-optimal across all states. Regret is well-defined in all settings, including cases with multiple treatments and when patients vary in observable covariates that may be used to differentiate treatment.

The regret concept is especially transparent when there are two treatments and one considers patients who are observationally identical, all having the same observed covariates. Let there be two treatments, A and B. In a state where A is better, the regret of a statistical treatment rule is the product of the probability across repeated samples that the rule commits a Type I error (choosing B) and the magnitude of the loss in expected welfare that occurs when choosing B. In a state where B is better, regret is the probability of a Type II error (choosing A) times the magnitude of the loss in expected welfare when choosing A.

Recall the earlier critique of the use of hypothesis testing to choose treatments. I called attention to the asymmetric treatment of Type I and Type II error probabilities and the inattention to magnitudes of welfare losses when errors occur. Evaluation of treatment rules by regret overcomes these problems. Regret considers Type I and II error probabilities symmetrically. In addition, it measures the magnitudes of the losses that errors produce.

### 3.5. Designing Trials to Enable Near-Optimal Treatment Choice

An ideal objective in designing trials would be to collect data that enable subsequent implementation of an optimal treatment rule in a patient population of interest; a rule that chooses the best treatment with zero chance of error. This is not achievable with finite sample size, but near-optimal rules---ones with small maximum regret---exist when trials are large enough.

Manski and Tetenov (2016) investigated trial design to enable near-optimal treatment choice. It was shown that, for any $\varepsilon > 0$, *ε-optimal* rules exist when sample size is large enough. An $\varepsilon$-optimal rule has expected welfare across repeated samples within $\varepsilon$ of the welfare of the best treatment, whatever the true state of nature may be.

Choosing sample size to yield $\varepsilon$-optimal treatment rules requires specification of a value for $\varepsilon$. The selected $\varepsilon$ determines how large a deviation from optimality a decision maker is willing to tolerate. In a medical context, it should be specified by clinical researchers concerned with patient care. Clinicians may perhaps want to set $\varepsilon$ equal to the *minimum clinically important difference* (MCID) in the average treatment effect comparing alternative treatments. Writers often call an average treatment effect clinically significant if its magnitude is greater than a value deemed minimally consequential in clinical practice. Computation of MCID or a related concept is already routine in clinical research because it must be performed when calculating statistical power in hypothesis testing.

Manski and Tetenov (2016, 2019, 2021) concluded that trial sample sizes set by clinically meaningful near-optimality criteria are much smaller than ones set to satisfy conventional statistical power criteria. Reduction of sample size can be beneficial in many ways. It lowers the cost of executing trials, the time needed to recruit subjects, and the complexity of managing multi-center trials.

*Illustration: Reconsidering Sample Size in the MSLT-II Trial*

Manski and Tetenov (2019) quantified the reduction in sample size in a particular setting, reconsidering a trial that compared aggressive treatment of patients with surveillance. The Multicenter Selective

Lymphadenectomy Trial II (MSLT-II) compared lymph node dissection and nodal observation for certain melanoma patients; see Faries *et al*. (2017). Using a statistical power calculation considering patient survival for a specified horizon to be the primary outcome, the MSLT-II investigators assigned 971 patients to dissection and 968 to observation.

We considered sample size from the perspective of near-optimal treatment choice. We assumed a simple patient welfare function that combines concern with patient survival and experience of the side effect of lymphedema. Welfare with nodal observation equal 1 if a patient survives for three years and equal 0 otherwise. Welfare with dissection depends on whether a patient experiences lymphedema. When a patient does not experience lymphedema, welfare with dissection equals 1 if the patient survives for three years and equals 0 otherwise. When a patient experiences lymphedema, welfare is lowered by a specified fraction h, whose value expresses the harm associated with lymphedema. Thus, a patient who experiences lymphedema has welfare $1 - h$ if he survives and $-h$ if he does not survive.

We determine a sample size that enables near-optimal treatment when $h = 0.2$ and $\varepsilon = 0.0085$. Setting $h = 0.2$ supposes that suffering from lymphedema reduces a patient's welfare by one-fifth relative to lymphedema-free survival. This quantification of the welfare loss produced by lymphedema was suggested by Cheville *et al*. (2010).

Setting $\varepsilon = 0.0085$ followed naturally from the way that the MSLT-II investigators performed their power calculation. They judged a difference of 5 percentage points in melanoma-specific survival to be a clinically meaningful loss in patient welfare and they judged 0.17 to be an acceptable probability of Type II error. Regret equals the magnitude of welfare loss times the probability that the loss will occur. Thus, the MSLT-II investigators judged $0.17 \times 0.05 = 0.0085$ to be an acceptable level of regret.

When $h = 0.2$ and $\varepsilon = 0.0085$, we found that near-optimal treatment is achievable if one sets $N = 488$ and assigns an equal number of patients to each treatment arm; that is, 244 to observation and 244 to dissection. This sample size is much smaller than the 1939 subjects who were enrolled in MSLT-II.

4. Evaluating Prediction Methods

Whereas much of Section 3 considered general problems of comparing treatments, I now specifically consider settings where empirical research seeks to learn probabilities of illness conditional on patient covariates, as in Section 2. In these contexts, a clinician ideally wants to predict illness well enough to make optimal treatment choices with certainty.

For simplicity, I focus on cases in which treatment B is better if a single illness probability is above a known threshold determined by the patient welfare function, with A better otherwise. That is, B is optimal if $P(y = 1|x) \geq t_x$ and sub-optimal if $P(y = 1|x) < t_x$, where $t_x$ is the threshold. Section 2 formalized two such settings in choice between surveillance and aggressive treatment. In one, aggressive treatment always prevents disease development. In the other, treatment does not affect disease development but affects the severity of illness when it occurs.

It is generally infeasible to optimize treatment with certainty. Clinical researchers are well-aware that statistical imprecision may degrade decisions when the available data are a finite sample of (illness, covariate, treatment) triples. Less well-appreciated but often more severe difficulties are identification problems, which are inferential difficulties that do not diminish as sample size increases.

Data quality may be imperfect due to missingness or mismeasurement of illness, covariates, or treatments. The *selection problem* occurs when observed treatments are generated not by known randomized mechanisms but by processes that are incompletely understood. Self-selection of treatments is the norm in observational data settings and in randomized trials where compliance with treatment assignment is not mandatory. Manski (2003, 2007), Tamer (2010), and Molinari (2020) exposit econometric research on these and other identification problems. A rare thoughtful analysis of identification of the selection problem in a clinical setting is Bhattacharya *et al.* (2012).

Statistical decision theory provides an appropriate framework to evaluate prediction methods in terms of how they inform clinical decisions; see Manski (2023) and Dominitz and Manski (2026a, 2026b). However, clinical researchers rarely use statistical decision theory and, when they do, they only use

Bayesian versions. The norm is to view prediction as a self-contained objective, not as a task undertaken to inform treatment choice. Sections 4.1 through 4.3 describe and critique prominent conventional practices.

### 4.1. Using Mean Square Error to Measure Predictive Performance

Viewing prediction as a self-contained objective, clinical research regularly uses the statistical concept of mean square error (MSE) to measure predictive performance with specified covariates. One considers the goal to be to learn the conditional probabilities $P(y = 1|x)$, $x \in X$. Here $y = 1$ is occurrence of an illness, x is a vector of observable patient covariates, and X is the set of all possible covariate values.

A standard formalization assumes that one observes a random sample of N realizations of (y, x) drawn from a population with distribution P(y, x). One specifies a model for $P(y = 1|x)$, say $F(x, \theta)$, where $\theta$ is a finite-dimensional parameter. A prevalent choice in clinical research has been a binary logit model, introduced by Berkson (1944). One uses the least squares or maximum likelihood method to compute a point estimate of $\theta$, say $\theta_N$.

Mean square error is $E_{\theta N}\{E_x[F(x, \theta_N) - P(y = 1|x)]^2\}$. The inner expectation is taken over the population distribution of x. The outer expectation is over the sampling distribution of $\theta_N$, which is determined by P(y, x), sample size, and the method used to estimate $\theta_N$. These distributions are not known and so must also be estimated using the sample data. With this done, researchers report an estimate of MSE.

From the perspective of clinical decisions, the central difficulty with this research practice is that the concept of MSE is disconnected from the objective of optimizing patient care. Section 2 pointed out that optimal choice between two treatments does not require precise knowledge of illness probabilities. It only requires determination of the correct inequality in equation (2).

A common clinical practice uses *plug-in* prediction to choose a treatment. That is, a clinician evaluates the inequalities in (2) using $F(x, \theta_N)$ to estimate $P(y = 1|x)$. The consequences for patient care depend on the relationship between $F(x, \theta_N)$ and $P(y = 1|x)$, but not as expressed by MSE. MSE measures the deviation between $F(x, \theta_N)$ and $P(y = 1|x)$ symmetrically for positive and negative discrepancies. What matters for patient care is (i) whether $F(x, \theta_N)$ and $P(y = 1|x)$ are on the same side of the decision threshold and (ii) the magnitude of regret (i. e., loss in expected welfare) when $F(x, \theta_N)$ and $P(y = 1|x)$ are on different sides of the decision threshold.

Heuristically, small MSE suggests that using $F(x, \theta_N)$ to estimate $P(y = 1|x)$ is inconsequential for patient care. However, large MSE does not suggest that plug-in prediction degrades care. The consequence of a deviation between $F(x, \theta_N)$ and $P(y = 1|x)$ depends on both the magnitude of the deviation and the sign. If $P(y = 1|x)$ is above the threshold for choice of treatment B, a positive deviation between $F(x, \theta_N)$ and $P(y = 1|x)$ does not affect the decision, regardless of the magnitude of the deviation. A negative deviation reduces a patient's expected welfare only if the magnitude is large enough to place $F(x, \theta_N)$ below the decision threshold. Analogous asymmetries occur if $P(y = 1|x)$ is below the decision threshold.

## 4.2. Using the Receiver Operating Curve to Measure Predictive Performance

Let us continue to consider clinical decision problems where treatment B is optimal if $P(y = 1|x) \geq t_x$ and sub-optimal if $P(y = 1|x) < t_x$, where $t_x$ is a specified threshold determined by the patient welfare function. Let $F(x, \theta_N)$ continue to be an estimate of $P(y = 1|x)$. For any x-invariant $\tau \in [0, 1]$, consider a binary classifier that predicts $y = 1$ if $F(x, \theta_N) > \tau$ and $y = 0$ if $F(x, \theta_N) < \tau$. Clinicians often convert probabilistic predictions of illness into binary ones in this manner.

A common practice in clinical research is to measure predictive performance using threshold $\tau$ by the true and false positive rates, defined as $P[F(x, \theta_N) > \tau)|y = 1]$ and $P[F(x, \theta_N) > \tau)|y = 0]$. Given a parameter estimate $\theta_N$, the probabilities P are taken over the covariate distributions $P(x|y = 1)$ and $P(x|y = 0)$. Both rates weakly decrease in $\tau$, so the true positive rate weakly increases with the false positive rate. The receiver operating curve (ROC) plots the true positive rate against the false positive rate, letting $\tau$ vary from 0 to 1. For example, see Bradley (1997).

Research studying the ROC uses the curve to provide a graphical expression of the tradeoff between the true and false positive rates that occur as the threshold varies. Each point on the curve is attainable for some value of $\tau$, so selecting a point on the curve is equivalent to selecting a threshold. They also use the ROC to provide a scalar summary of prediction performance, one that is not threshold dependent. It is common to measure predictive performance by the area under the curve (AUC).

I am not aware of any use of the ROC that coherently informs clinical decisions. The true and false positive rates, as defined in the ROC, are not relevant to treatment choice. The relevant probability of illness is $P(y = 1|x)$. True and false positive rates related to $P(y = 1|x)$ would be $P[y = 1|F(x, \theta_N) > \tau]$ and $P[y = 0|F(x, \theta_N) > \tau]$, not the ones used in the ROC. Nevertheless, study of the ROC has been prominent in clinical research.

### 4.3. Measuring the Value of Covariate Information

Sections 4.1 and 4.2 considered measurement of predictive performance with specified patient covariates. A literature in clinical research seeks to measure the value of observing additional covariates beyond those initially observed. This matter is particularly important when a clinician contemplates whether to order diagnostic tests for patients, which may be costly and/or invasive.

#### 4.3.1. Decision Theoretic Assessment

Assessment of the value of additional covariate information from the perspective of clinical decisions was apparently first studied by Pauker and Kassirer (1980) and Phelps and Mushlin (1988). More recently, see Basu and Meltzer (2007), Manski (2013), Cassidy and Manski (2019), and Manski, Mullahy, and Venkataramani (2023). The basic idea is simple.

With the decision model of Section 2, optimal expected welfare conditioning on covariates x is

$$\max[P(y = 1|x)\cdot u_x(1, A) + P(y = 0|x)\cdot u_x(0, A),\ P(y = 1|x)\cdot u_x(1, B) + P(y = 0|x)\cdot u_x(0, B)]. \tag{7}$$

Let an additional covariate w be observed for each patient and let treatment choices be made conditional on (x, w) rather than x. Suppose that illness probabilities may vary with w, but not patient welfare in a given illness status. Then optimal expected welfare for patients with covariates x is

$$(8) \quad E_{w|x}\{\max[P(y = 1|x, w)\cdot u_x(1, A) + P(y = 0|x, w)\cdot u_x(0, A),$$
$$P(y = 1|x, w)\cdot u_x(1, B) + P(y = 1|x, w)\cdot u_x(0, B)]\}.$$

Here $E_{w|x}$ is the expected value with respect to the distribution P(w|x). Quantity (8) is greater than or equal to (7) by Jensen's Inequality. Hence, observation of w weakly increases optimal expected welfare. Manski, Mullahy, and Venkataramani (2023) quantify the magnitude of the increase.

4.3.2. Predictive Value and Sensitivity/Specificity

Viewing prediction as a self-contained objective, clinical research has not embraced the above decision theoretic way to measure the value of covariate information. Instead, it has applied two sets of connected but distinct concepts: positive/negative predictive value and sensitivity/specificity.

The clinical literature has mainly considered settings where the additional covariate is binary, with w = 1 suggesting high risk of illness and w = 0 low risk. A leading case occurs when w is the finding of a diagnostic test whose outcome is communicated to clinicians in a binary manner. Then w = 1 and w = 0 are termed positive and negative test findings. I focus on this case, drawing on Manski (2021b).

Some clinical research compares the conditional illness probabilities P(y|x, w = 1) and P(y|x, w = 0). Performing a test is deemed to have no value if P(y|x, w = 1) = P(y|x, w = 0) and to increase in value as the distance between these distributions grows. The probabilities P(y = 1|x, w = 1) and P(y = 0|x, w = 0) are called positive and negative predictive values (PPV and NPV), whereas P(y = 1|x) is called the prevalence

of illness. Researchers measure the information value of a test by the absolute risk $P(y = 1|x, w = 1) - P(y = 1|x, w = 0)$ or the relative risk $P(y = 1|x, w = 1)/P(y = 1|x, w = 0)$.[6]

Much clinical research measures the value of tests in a different way, focusing on sensitivity and specificity (Yerushalmy, 1947). Sensitivity is the probability $P(w = 1|x, y = 1)$ that an ill person receives a positive test finding. Specificity is the probability $P(w = 0|x, y = 0)$ that a non-ill person receives a negative finding. Thus, sensitivity and specificity are concerned with prediction of the test finding given knowledge of a patient's illness status. These quantities are mathematically similar to those studied in ROC analysis.

Sensitivity and specificity do not inform clinical decisions. When making a treatment choice, a clinician observes a test finding and wants to predict whether a person is ill. The clinician does not observe whether a person is ill and want to predict a test finding. Altman and Bland (1994) recognized the issue, writing (p. 102):

> "The whole point of a diagnostic test is to use it to make a diagnosis, so we need to know the probability that the test will give the correct diagnosis. The sensitivity and specificity do not give us this information. Instead we must approach the data from the direction of the test results, using predictive values. Positive predictive value is the proportion of patients with positive test results who are correctly diagnosed. Negative predictive value is the proportion of patients with negative test results who are correctly diagnosed."

Also see Feinstein (2002).

Given that PPV and NPV are clinically relevant concepts while sensitivity and specificity are not, it is important to ask why measurement of test value often focuses on the latter quantities rather than the former. Part of the answer appears to be that researchers find it easier to measure sensitivity and specificity.

---

[6] Clinical research has emphasized relative rather than absolute risk. This is hard to justify from the perspective of clinical decisions. For example, consider two scenarios. In one, the probability of illness conditional on $w = 1$ is 0.12 and conditional on $w = 0$ is 0.08. In the other, these probabilities are 0.00012 and 0.00008. Relative risk in both scenarios is 1.5. Absolute risk is 0.04 in the first scenario and 0.00004 in the second. The first scenario is clearly much more clinically relevant than the second. Relative risk does not differentiate the scenarios, but absolute risk does.

Measurement by relative rather than absolute risk has long been criticized; see Berkson (1958), Fleiss (1981, Section 6.3), and Hsieh, Manski, and McFadden (1985). The rationale, such as it is, rests on the widespread use of retrospective rather than random sampling of patient populations in clinical research studying observational data. Cornfeld (1951) showed that the data generated by retrospective sampling point-identify relative risk under the rare-disease assumption. These data only partially identify absolute risk. Fleiss (1981) remarked that (p. 92) "retrospective studies are incapable of providing estimates" of absolute risk. Manski (2007, Chapter 5) proved that such studies informatively bound absolute risk.

Estimation of PPV and NPV requires observation of test results and illness status for a representative sample of the relevant population, say through random sampling. It often is easy to observe test results but not illness status. Indeed, if it were easy to observe illness status, tests would serve no practical purpose.

Whereas observation of illness status in a patient population of interest may be difficult, researchers sometimes are able to do so for special groups of patients. This enables estimation of sensitivity and specificity for these groups. The practice has been to estimate in this manner and assume that the findings obtained for the special groups also hold in the relevant population.

Suppose that sensitivity and specificity have been estimated in some special groups and that one believes it credible to extrapolate to the relevant patient population. Then PPV and NPV can be derived via Bayes Theorem if one additionally knows the prevalence of the disease in this population. However, it is often difficult to measure prevalence precisely.

Aiming to cope constructively with this problem, Manski (2021b) studied partial identification of predictive values given bounds on prevalence. Partial identification analysis removes the traditional focus on point estimation obtained under strong assumptions. Instead it uses weak assumptions that aim to be credible in the context under consideration. The results are credible bounds on predictive values. I used testing for COVID-19 to illustrate. See Obradovic (2025) for further partial identification analysis.

## 5. Study-Centered and Patient-Centered Meta-Analysis

The problematic conventions of clinical research discussed above concern analysis of findings from single studies. A further poorly motivated convention appears when researchers attempt to combine findings from multiple studies. The impetus to combine findings is understandable. Clinicians want to interpret the mass of information provided by evidence-based research. The hard question is how to interpret this information sensibly.

Combination of findings has long been performed by narrative and systematic review of a set of studies. Concerned that such reviews are subjective, statisticians proposed *meta-analysis* in an effort to provide an objective methodology for combining the findings of multiple studies. Clinical research has valued meta-analysis to the extent that it is commonly placed at the top level of a so-called "evidence pyramid" (e.g., Murad *et al.*, 2016). My discussion draws on Manski (2020a).

Meta-analysis was originally developed to address a purely statistical problem. Suppose that one wants to estimate as well as possible some parameter characterizing a study population. For example, the parameter may be the mean health outcome that would occur if all patients in a population were to receive a specified treatment. Suppose that multiple studies have been performed in this population, each drawing a random sample from the population and assigning treatments randomly. If the raw data on the outcomes are available, the most precise way to estimate the parameter is to combine the samples and compute the estimate using all the data. However, the raw data often are unavailable, making it infeasible to combine the samples. Instead, let multiple parameter estimates be available, each computed with the data from a different sample. Then meta-analysis proposes methods to combine the multiple estimates. The usual proposal is to compute a weighted-average of the estimates, the weights varying with sample size. See Hedges and Olkin (1985).

The original concept of meta-analysis described above is uncontroversial, but its applicability is limited. It is rare that multiple studies are performed on the same population. It is common for multiple trials to be performed on distinct patient populations that may have different distributions of treatment response. The process for selection of treatments and the measurement of outcomes may vary across studies as well.

Clinical research regularly performs meta-analysis in such settings. Gene Glass, who introduced the term *meta-analysis*, recognized the challenge of combining findings from disparate studies. He wrote (Glass, 1977, p. 358): "The tough intellectual work in many applied fields is to make incommensurables commensurable, in short, to compare apples and oranges."

### 5.1. Study-Centered Meta-Analysis

To operationalize comparison of "apples and oranges," it has become common in clinical research to perform meta-analyses that use a random-effects model proposed by DerSimonian and Laird (1986). This model recognizes that each study examined in a meta-analysis may concern a distinct value of a parameter of interest. It places structure on the heterogeneity of parameters across studies by supposing that each study is drawn at random from what DerSimonian and Laird called "a population of potential studies" (p. 181). They interpreted a weighted average of the estimates across studies as an estimate of the mean parameter value across studies. They used the variance of the parameters to measure the extent to which the parameter value varies across studies.

The mathematics of this random-effects model is transparent, but its relevance to clinical decisions is obscure. DerSimonian and Laird did not explain what is meant by "a population of possible studies," nor why the studies being combined should be considered a random sample from this population. They did not explain how a mean outcome across a population of possible studies connects to what might matter to a clinician, namely the mean outcome across a population of patients.

Clinical researchers who perform meta-analysis using the DerSimonian and Laird random-effects model struggle to explain coherently how clinicians should use the findings. Consider, for example, the meta-analysis performed by Chen and Parmigiani (2007) of ten disparate studies predicting risk of breast and ovarian cancer among women who carry BRCA mutations. The authors described a weighted average of the risks reported by all studies as a (p. 1331) "consensus estimate." However, there was no consensus across studies, which reported heterogeneous estimates pertaining to heterogeneous populations.

### 5.2. Patient-Centered Meta-Analysis

Manski (2020a) studied meta-analysis as a problem in partial identification. Suppose that M disparate studies have been performed. Let the objective be to combine their findings to learn about a patient population of interest. In particular, clinicians may want to learn the probability of illness conditional on specified covariates.

I pointed out that, except in rare cases where at least one study is performed on the exact population of interest and data quality is perfect, a researcher making credible assumptions will assess that each study partially rather than point identifies the population parameter of interest. That is, it is often credible to think that, even in the absence of statistical imprecision, each study yields some information about the parameter, but not its precise value. A feasible value of the parameter is consistent with everything learned in the M studies if and only if it lies within each of the M identification regions. Thus, the identification region combining the studies is the intersection of the regions obtained with each study.

Set intersection is easy to perform when the parameter is real-valued and the M identification regions are intervals. Then the intersection is itself an interval, whose lower bound is the greatest lower bound of the study-specific intervals and whose upper bound is the least upper bound of these intervals. For example, let three studies of patient survival be available. Suppose that their identification regions for the probability of survival are [0.4, 0.7], [0.2, 0.6], and [0.5, 1]. Set intersection yields [0.5, 0.6] as the bound on survival probability obtained by combining the studies.[7]

## 6. Prediction with Incredible Certitude

The systemic methodological dysfunction discussed in Sections 3 through 5 stems from conceptual deficiencies that distance clinical research from clinical decisions. Further deficiencies result from prediction with incredible certitude (Manski, 2011, 2020b). This occurs when researchers make strong unsubstantiated assumptions about the functional form of conditional illness probabilities and the quality of available data. These assumptions may or may not be correct. If not, estimated illness probabilities $F(x, \theta_N)$ may not converge to actual probabilities $P(y = 1|x)$ as sample size increases. Treatment choices based on incorrect estimated illness probabilities may be sub-optimal.

---

[7] This example demonstrates a subtlety in the operation of set intersection. One might intuit that when M studies yield identification regions of different sizes, the studies yielding the smallest identification regions would play the most prominent role when combining findings. The example shows that this need not be the case. The lengths of the three study-specific intervals are 0.3, 0.4, and 0.5 respectively. Yet the set intersection is determined entirely by the second and third intervals. The lesson is that the result of set intersection depends on the joint positioning of the sets, not only by their sizes.

Section 6.1 calls attention to incredible certitude in conventional specification of prediction models. Section 6.2 cautions against wishful extrapolation from trials to patient populations of clinical concern. Section 6.3 extends the caution to wishful prediction with missing data.

## 6.1. Conventional Prediction Models

Clinical researchers rarely have good reasons to believe that illness probabilities vary with covariates in the manner assumed by conventional parametric models. Nevertheless, they apply these models routinely.

When viewing illness as a binary outcome, the convention has been to assume the logit model $P(y = 1|x) = e^{x\theta}/(1 + e^{x\theta})$ of Berkson (1944). The parameters are estimated by maximum likelihood or least squares. Statistical imprecision is assessed using asymptotic statistical theory, assuming the model is correctly specified.

When analyzing longitudinal data on the onset of illness, the convention has been to assume the proportional hazards survival model of Cox (1972). The hazard rate for illness onset at date T is assumed to be $h(T|x) = h_0(T)\cdot e^{x\theta}$, where $h_0(T)$ is the time-varying baseline hazard and $e^{x\theta}$ is the x-varying relative risk. The probability of the binary event that illness onset occurs on or before a specified date is derived from the hazard function. The baseline hazard is estimated nonparametrically and $\theta$ by Cox's partial likelihood method. Again, statistical imprecision is assessed using asymptotic theory, assuming the model is correctly specified.

The logit and proportional hazards model make strong assumptions that may or may not be realistic. Nevertheless, clinical researchers interpret model predictions of illness probabilities as if they are correct.

Indeed, reviewers and editors of research papers submitted to leading medical journals expect that these models will be used rather than others that make alternative parametric assumptions. Use of more credible nonparametric statistical methods to predict illness conditional on patient covariates has been rare in clinical research.[8]

6.2. Wishful Extrapolation of Treatment Response from Trials to Clinical Populations

Clinical researchers often use untenable assumptions to extrapolate predictions from study populations to clinical populations. I have called this manifestation of incredible certitude *wishful extrapolation*. Manski (2019a) critiqued extrapolation from randomized trials. I summarize here.

Trials have long enjoyed a favored status within clinical research, often being called the 'gold standard' for such research. The appeal of trials is that, with sufficient sample size and complete observation of outcomes, they deliver credible findings on treatment response in the study population. However, extrapolation of findings from trials to clinical treatment choice can be difficult. Researchers and clinicians use untenable assumptions to extrapolate.

One substantial problem is that study populations in trials often differ from actual patient populations. It is common to perform trials studying treatment of a specific disease only on subjects who have no co-morbidities. Moreover, study populations consist of individuals with specified demographic attributes who volunteer to participate in a trial. Participation in a trial may be restricted to individuals in certain age

---

[8] Although use of nonparametric statistical theory has been rare in clinical research, there appears to be a growing willingness of clinicians to use nonparametric illness predictions generated by computer scientists who use the terms *machine learning (ML)* and *artificial* intelligence (AI) to describe their algorithms. Computer scientists do not use statistical theory to evaluate the performance of their predictions. Instead, they use heuristically motivated "out-of-sample" (OOS) evaluation approaches. See the discussion in Dominitz and Manski (2026a).

It is intriguing that clinicians tightly adhere to statistical theory when performing hypothesis tests and measuring the imprecision of parameter estimates, yet they are increasingly willing to embrace ML-AI predictions that lack any foundation in statistical theory. I conjecture that this dissonance is explained by a profound cultural difference between biostatisticians and computer scientists. Biostatisticians have been highly conservative in their adherence to statistical theory. Computer scientists optimistically assert that their algorithms "work in practice" even though they are "black boxes." Such assertions may be enticing to some clinicians who have used statistical theory without understanding it.

categories who reside in certain locales. Volunteers are those who respond to financial and medical incentives to participate. It often is wishful extrapolation to assume that treatment response in trials performed on volunteers with specified demographic attributes who lack co-morbidities is the same as in clinical populations. Basu and Gujral (2020) is a rare analysis studying the consequence of self-selection of subjects into trials.

A further problem is that treatments in trials may differ from those administered in clinical practice. This is particularly so in trials comparing drug treatments. Drug trials are commonly double-blinded, neither the patient nor the clinician knowing the assigned treatment. A double-blinded drug trial reveals the distribution of response in a setting where patients and clinicians are uncertain what treatment a patient is receiving. It does not reveal what response would be when patients and clinicians know what drug is being administered and can react to this information.

For example, consider drug treatments for hypertension. Patients may react heterogeneously to the various drugs available for prescription. A clinician treating a specific patient may sequentially prescribe alternative drugs, trying each for a period in an effort to find one that performs satisfactorily. Sequential experimentation is not possible in a blinded trial. The standard protocol prohibits the clinician from knowing what drug a subject is receiving and from using judgment to modify the treatment. Blinding is also problematic for interpretation of noncompliance with assigned treatments in trials.

### 6.3. Wishful Prediction with Missing Data

Extrapolation is an extreme case of prediction with missing data. One wants to predict illness in a population of interest, but one does not observe this population. Less dramatic but still vexing cases of prediction with missing data occur when one draws sample data from the population of interest, but some realizations of outcomes and/or covariates are not observed. These missing data problems arise regularly in clinical research.

When confronting missing data, an analog to wishful extrapolation is the conventional practice of assuming that data are missing at random (MAR). Let $z = 1$ when a realized (outcome, covariate) pair is observed and $z = 0$ when some part of the pair is not observed. A commonly used version of the MAR assumption is that $P(y|x, z = 0) = P(y|x, z = 1)$. Maintaining this assumption, researchers often study illness in the sub-population with $z = 1$ and wishfully extrapolate findings to the sub-population with $z = 0$, thus yielding predictions for the entire patient population. Other versions of the MAR assumption may be used when either outcome or covariate data are missing, but not both simultaneously. In these settings, MAR assumptions are used to motivate imputation of missing data. Imputation uses observed data to predict the values of missing data and then handles the predictions as if they are actual observations.

The severe problem with MAR assumptions is that they rarely have a credible basis. The difficulty is akin to the selection problem discussed in Section 4. That is, missingness may not actually be random but rather be generated by processes that are incompletely understood. Clinical researchers should be aware of this, yet MAR conventions are so strong that these assumptions are rarely questioned. Writing on this phenomenon recently, I called it "the allure of making stuff up" (Manski, 2025b). Clinical research hardly ever makes use of the extensive econometric research on credible partial identification with missing data, reviewed in Manski (2003, 2007) and elsewhere.

## 7. Towards Healthy Clinical Research

I wrote in the Introduction that the systemic dysfunction in statistical clinical research is held in place by three factors: (1) the rudimentary instruction in statistical methodology received by medical students and residents; (2) the reliance of clinical researchers on consulting biostatisticians who act as gatekeepers in evaluation of grant proposals and paper submissions, approving research practices that adhere to convention and rejecting nonstandard practices; and (3) the institutional practices of research funding agencies, medical journals, and governmental bodies, which make adherence to convention a prerequisite for award of grants, acceptance of paper submissions, and approval of treatment innovations.

The unfortunate consequence of the systemic dysfunction is that clinical research yields inadequate evidence interpreted by inappropriate methods. This situation should be distressing to patients, as it yields sub-optimal care. It is also intellectually curious. Clinical medicine has benefitted greatly from research in genomics, improvements in diagnostic and surgical tools, and discovery of new drug treatments. Yet statistical practices have remained stuck in place. Using economic terminology, I see the continuing dysfunction as an inferior equilibrium. Statistical clinical research remains in a harmful stable state which is difficult to escape because individual participants lack the incentive to change their behavior.

Movement from an inferior to a better equilibrium may occur if a subset of influential participants in a system recognize the problem and collectively change their behavior. When I wrote Manski (2019a), I optimistically thought that medical schools would act constructively if they were to become aware of the problem. I concluded the main text as follows (p. 119):

> "It would be useful to introduce medical students to core concepts of uncertainty quantification (both statistical imprecision and identification) and to decision analysis as part of their basic education. However, I do not expect that this will suffice. The basic training of clinicians already covers a considerable array of subjects. Substantial additions to curricula may be difficult to achieve.
>
> I expect that an important part of the solution will be to bring specialists in uncertainty quantification and decision analysis into the clinical team. Modern clinical practice commonly has a group of professionals jointly contribute to patient care. Surgeons and internists may work together, and in conjunction with nurses and technical personnel. However, existing patient-care teams do not ordinarily draw on professionals having specific expertise in the framing and analysis of complex decision problems. . .
>
> To instruct basic medical students and develop specialists in patient care under uncertainty will require that medical schools create and implement appropriate curricula. I urge medical educators to make it so."

I now think I was naïve to expect that medical schools will take the lead on their own. Medical educators know that, after completing their training, young clinical researchers will suffer if they do not adhere to the conventions mandated by funding agencies such as NIH, medical journals, and governmental bodies such as the FDA. As I now see it, meaningful change can occur if the leadership of an influential subset of these institutions and of medical schools take the initiative. If not, the present systemic dysfunction will continue.